\documentclass[aps,pre,showpacs,amsmath,amssymb,amsfonts,lengthcheck,twocolumn,longbibliography,superscriptaddress]{revtex4-2}

\usepackage[T1]{fontenc}
\usepackage[utf8]{inputenc}
\usepackage{amsmath,amssymb,amsfonts,bm}
\usepackage{amscd}
\usepackage{mathrsfs}

\usepackage{lipsum}
\usepackage{fmtcount}
\usepackage{chemformula}
\usepackage{dutchcal}
\DeclareMathAlphabet{\matholdcal}{OMS}{cmsy}{m}{n}

\usepackage[bookmarks=true,colorlinks,linkcolor=blue,urlcolor=blue,citecolor=blue]{hyperref}

\usepackage{bbm}

\newcommand{\bla}{bla\\bla\\bla\\bla\\bla}

\usepackage{fmtcount}

\newcommand{\be}{\begin{equation}}
\newcommand{\ee}{\end{equation}}
\newcommand{\bea}{\begin{eqnarray}}
\newcommand{\eea}{\end{eqnarray}}

\begin{document}

\title{Maxwell's demon with dementia -- the deterioration of information transcription}
\author{Maggie Williams}
\email{mwilliams@umbc.edu}
\affiliation{Department of Physics, University of Maryland, Baltimore County, Baltimore, MD 21250, USA}
\affiliation{Quantum Science Institute, University of Maryland, Baltimore County, Baltimore, MD 21250, USA}
\author{Emery Doucet}
\affiliation{Department of Physics, University of Massachusetts, Boston, Massachusetts 02125, USA}
\affiliation{Department of Physics, University of Maryland, Baltimore County, Baltimore, MD 21250, USA}
\affiliation{Quantum Science Institute, University of Maryland, Baltimore County, Baltimore, MD 21250, USA}
\author{Sebastian Deffner}
\affiliation{Department of Physics, University of Maryland, Baltimore County, Baltimore, MD 21250, USA}
\affiliation{Quantum Science Institute, University of Maryland, Baltimore County, Baltimore, MD 21250, USA}
\affiliation{National Quantum Laboratory, College Park, MD 20740, USA}
\date{\today}

\begin{abstract}
In introductory biology, aging is typically explained as a result of mutations during the DNA replication process within cells. Upon abstraction, we recognize that cellular aging can be understood as the gradual decay in fidelity of information transcription. Since cellular processes are microscopic and inherently stochastic, the abstracted process of information transcription can be understood using Markovian dynamics. In our work, we model the process of information transcription with an autonomous Maxwell’s demon (AMD) which interacts with two bitstreams, a lifted mass, and a heat reservoir. As main results, we analyze the steady-state properties of the system with both time-independent and time-dependent transition rates, focusing on the statistics of extractable work, bit transcription fidelity, and two–bit mutual information. Together, these results provide a holistic view of a simplified model for DNA transcription as an information–theoretic process.
\end{abstract}

\maketitle

\section{Introduction}

Information processing plays a crucial part in the functioning of biological systems. From the replication of DNA to the synthesis of proteins, biological functions depend on the accurate transcription and translation of molecular sequences. Yet, these processes are not perfect: over time, cells accumulate transcriptional and translational errors that gradually erode their ability to maintain homeostasis. At a fundamental level, biological aging can therefore be understood as the progressive degradation of information-processing capacity in a stochastic, thermodynamically constrained system \cite{NoisyCellGrowth, quarton2020uncoupling}. Recent studies have shown that transcription errors in RNA can promote the formation of amyloid-like proteins, which are misfolded proteins that form fibrous aggregates and disrupt normal cellular function \cite{sipe2000amyloid, Amyloids, Gertz2015, Moreau2012, Dember2006, Schroder2013} in a variety of cell types \cite{Chung2023, transerror}. Since RNA is the informational channel by which the RNA polymerase transcribes complementary base pairs to the DNA \cite{Kornberg}, such transcriptional errors compromise protein synthesis and degrade cellular function \cite{Vermulst2015}.

Biological transcription and translation occur in environments dominated by thermal noise, where molecular interactions are inherently probabilistic. 
The dynamics of such systems are then naturally described by \textit{time-inhomogeneous}, Markov processes, in which transition rates evolve in response to internal or environmental perturbations \cite{peliti2021stochastic}. From this perspective, we can then model complex biological phenomena such as adaptation, fatigue, and aging, by encoding changes in biological function into time-dependent rate matrices. In particular, by allowing the rates that govern state transitions to vary smoothly in time, one can capture how a system’s ability to process and store information deteriorates over time. To formalize this connection, we model transcriptional fidelity within the framework of stochastic thermodynamics \cite{peliti2021stochastic} using a minimal model of an autonomous Maxwell’s demon \cite{maxwell}. 

To provide context, Maxwell’s demon was originally conceived as a thought experiment to test the limits of the second law of thermodynamics \cite{maxwell1871,szilard}. It was later recognized as a prototype of an information-processing system, laying the foundation for the thermodynamics of computation and measurement \cite{landauer,bennett}. Modern formulations reinterpret the demon as an \textit{autonomous information ratchet}, which is a system that converts input bitstreams into ordered outputs while coupled to thermal and work reservoirs \cite{maxwell,Deffner2013PRX,Boyd2016,Boyd2017PRE,Boyd2018PRE}. Recent work has shown that demons can extract work not only from biased inputs but also from \textit{correlated} environments, providing a direct link between information structure and thermodynamic capacity \cite{Horowitz2014JSM, Safranek2018PRA,Boyd2020, Boyd2022}. Likewise, continuous formulations of information flow in nonequilibrium systems demonstrate that mutual-information rates modify the second law, establishing general bounds on the work and entropy balance in feedback-driven or autonomous devices \cite{Horowitz2014PRX, Sagawa2012PRL, Prokopenko2015PRE,Sone2021entropy}. These systems quantify how information flow, energy dissipation, and entropy production determine performance, and their results have broadened Maxwell’s original thought experiment into a general framework for stochastic information processing described by Markovian dynamics. There are also many works that discuss how to characterize steady-state solutions to Markov processes with time-inhomogeneous transition rates \cite{parrondo_2023,stopnitzky_2019,dynamic_2024,chittari_2024,vanDijk1992,Arns2010InhomCTMC,Hsu2017NHCTMC,Lencastre2016,Ding2021,Truquet2019,Talkner2003,TalknerLuczka2005,Forastiere2022,RaoEsposito2018,Aslyamov2024,MoranLedezma2024}, but not as many sources that discuss this potential application to biological processes whose functionalities are prone to environmental fluctuations \cite{Innocentini2018,JiaLi2023,Wu2024Spectral,Fintzi2017,WuMGDrivE22021,Meng2022HMMCancer,Qian2006OpenNESS,Esquivel2025CTMCReview}.

Within this broader picture, molecular systems can be viewed as physical realizations of autonomous demons: RNA polymerase, for instance, operates as a finite-memory information ratchet that converts thermal fluctuations from its environment into ordered DNA transcription. The autonomous Maxwell’s demon model developed in the present paper builds directly on this foundation, providing a simplified, exactly solvable representation of information transcription under time-dependent driving. Our model provides a recipe, enabling the quantitative analysis of entropy production, work extraction, and mutual information of a Maxwell's demon system under time-dependent driving, discussing how its informational and energetic capabilities decay as the system transitions from a functional to a defective regime.

\section{Biologically Motivated Model}

\begin{figure}
    \centering
    \includegraphics[width=0.48\textwidth]{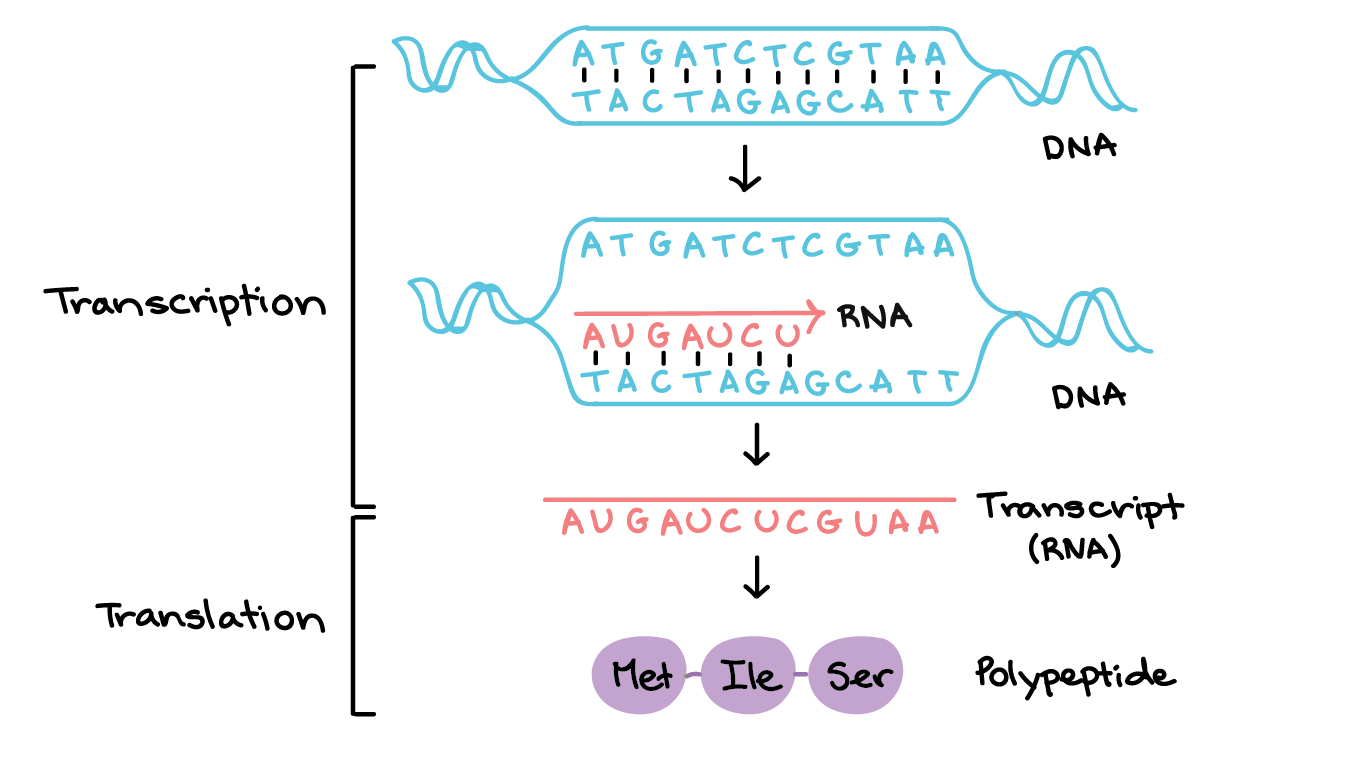}
    \caption{Illustration of DNA transcription, adapted from Khan Academy, \textit{“Overview of transcription”} (\href{https://bit.ly/4pXt3nY}{Khan Academy}), licensed under \textit{CC BY-NC-SA 4.0}.}
    \label{fig:placeholder}
\end{figure}

DNA transcription can be abstracted as an information-theoretic process in which the genetic information encoded in DNA is read and copied into messenger RNA by RNA polymerase, converting a physical molecular sequence into a transmitted sequence of information \cite{Watson1953, Ochoa1955, Kornberg1956, Kornberg2001, Levene1909}. This process is schematically depicted in Fig. \ref{fig:placeholder}. More abstractly, the process can be represented as the transcription of information, which has been modeled in the literature as a so-called \emph{Maxwellian information ratchet} \cite{Serreli2007,Spinney2018,Jurgens2020,Boyd2016}. 

Information ratchets are physical systems that convert thermal fluctuations into directed motion or ordered information flow, operating analogously to Maxwell’s demon. In such systems, microscopic degrees of freedom are exploited to extract useful work or to increase informational order, typically by coupling to a bitstream that encodes the system’s state transitions \cite{maxwell}. For our present purposes, the transcription process can be understood through the lens of an autonomous Maxwell’s demon: the RNA polymerase acts as an information-processing device that reads, interprets, and copies a sequence of molecular ``bits'', converting random thermal fluctuations into the ordered replication of DNA. 

Thus, DNA transcription can be modeled as an information-theoretic process, where the polymerase functions as an autonomous Maxwellian information ratchet that maintains fidelity and directionality under thermal noise. This abstraction allows the dynamics of information transcription to be analyzed in terms of stochastic thermodynamics and information flow.

\subsection{Time-Homogeneous Dynamics}

To construct the simplest possible model of information transcription, we encode the four nucleotide bases $(A, T, G, C)$ in a reduced binary form, mapping them onto a bit-wise basis of 0s and 1s. 

The autonomous demon interacts with three subsystems: a thermal reservoir of temperature $k_BT$, a suspended mass $m$, and two bitstreams. The suspended mass serves as a simplified representation for the mechanism of ATP synthase \cite{Boyer1997}. The demon is modeled as a Markov process with three internal states, consistent with the previous formulation in Ref. \cite{maxwell}, though in principle the state space may be generalized to an arbitrary number $n \geq 2$ of internal states. The demon makes transitions between states \{$\mathrm{A}, \mathrm{B}, \mathrm{C}$\} at fixed rates $R_{ij}$, enabling it to either extract work from the mass and perform information transcription or perform information erasure and release energy to the mass, depending on the direction of its transitions.

\begin{figure}
        \centering
        \includegraphics[width=.48\textwidth]{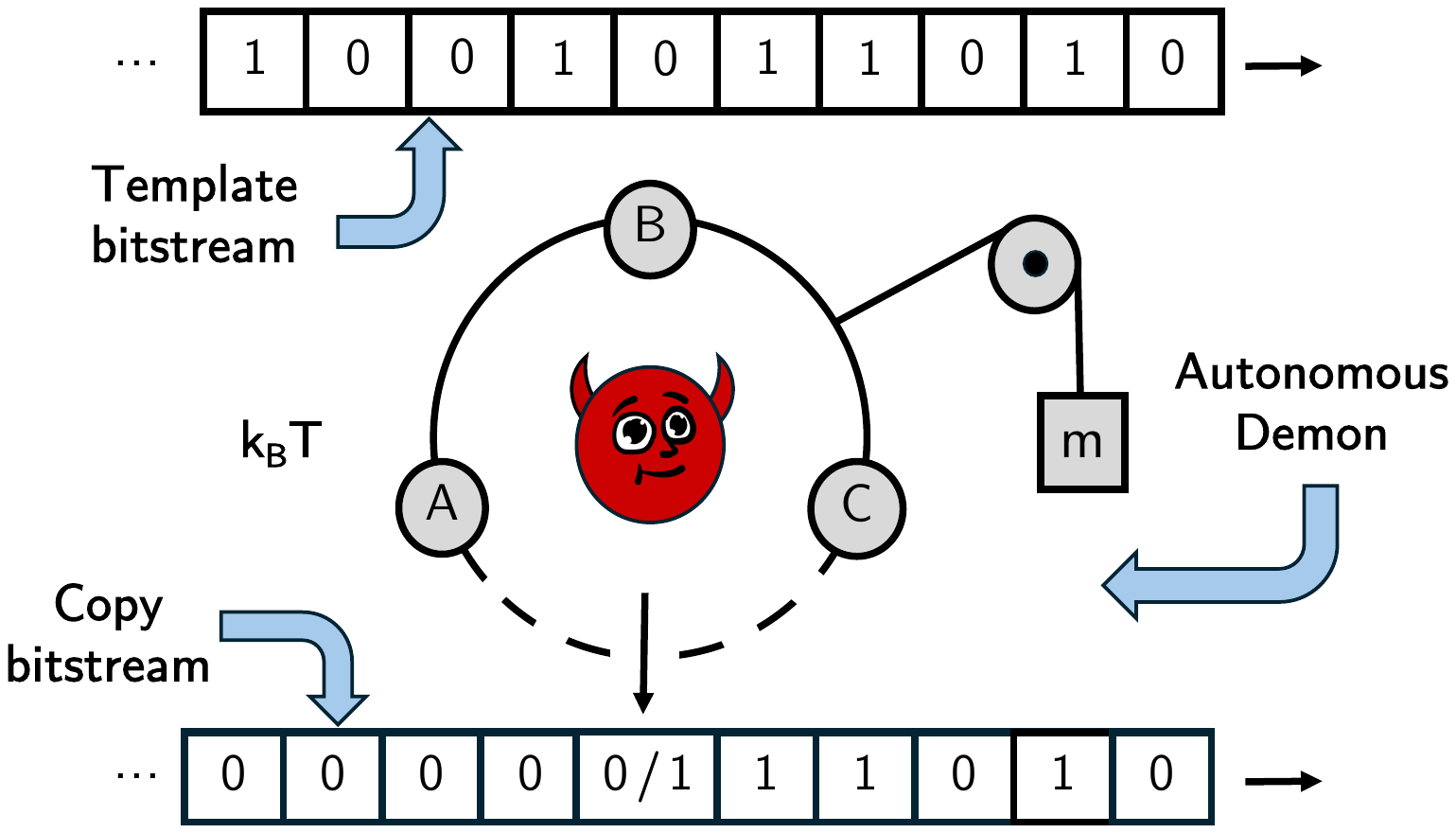}
        \caption{Autonomous demon system to simulate information transcription.}
        \label{fig:dna-transcription}
    \end{figure}

The `template' bitstream, representing an unzipped strand of DNA, is a random sequence of 0s and 1s. The autonomous demon plays the role of the RNA polymerase and reconstructs the `template' bits and transcribes them to a `copy' bitstream initialized to all 0s, representing the mRNA molecule. The model is depicted in Fig. \ref{fig:dna-transcription}. 

The dynamics of the composite demon-bit system are governed by the following transition rate matrix $\textbf{R}$,
\begin{equation}
\textbf{R} = \begin{pmatrix}
-1 & 1 & 0 & 0 & 0 & 0 \\
1 & -2 & 1 & 0 & 0 & 0 \\
0 & 1 & -2+\epsilon & 1+\epsilon & 0 & 0 \\
0 & 0 & 1-\epsilon & -2-\epsilon & 1 & 0 \\
0 & 0 & 0 & 1 & -2 & 1 \\
0 & 0 & 0 & 0 & 1 & -1
\label{eqn:rate-matrix}
\end{pmatrix}
\end{equation}
where $\epsilon$ represents the coupling strength to the corresponding bitstream. In App.~\ref{sec:appendix} we outline an analysis of the `read-write' fidelity, which justifies the use the identical value for $\epsilon$ for both bit streams.

Accordingly, the master equation governs the time evolution of the system,
\begin{equation}
    \label{eqn:master-equation}
    \dot{\textbf{P}} = \textbf{R} \, \textbf{P}(t),
\end{equation}
where $\textbf{P}(t)$ is the state probability vector. For the time-homogeneous case, the system is guaranteed to relax to a periodic steady-state distribution $\textbf{P}_\mathrm{{PSS}}$ \cite{vanKampen2007}, which satisfies,
\begin{equation}
\textbf{R} \, \textbf{P}_{\mathrm{PSS}} = 0.
\end{equation}
where $\textbf{P}_{\mathrm{PSS}}$ is the periodic steady-state probability vector. Here the steady state is periodic since the `template-copy' bit pair is reinitialized after once the interaction time, $\tau$, has passed. This can be conceptualized as the demon stopping at one bit pair, reading the `template' bit, and transcribing it into the `copy' register, and then moving onto the next bit pair after time $\tau$. 

As the demon reads a `template' bit, it flips the `copy' bit to the correct value to a copy of the original. For a perfect information transcription process, the 'copy' bit should exactly match that of the 'template' bit. 

For the following analysis, we solve the stationary-state dynamics for the $12\times12$ block diagonal matrix with the $6\times6$ rate matrix $\textbf{R}$ on the diagonal, and all other entries being zero. In the upper left quadrant, $\epsilon=-1$, corresponding to a perfect erasure $1 \rightarrow 0$. In the lower right quadrant, $\epsilon=+1$, corresponding to a perfect anti-erasure $0 \rightarrow 1$.

\begin{figure}
    \centering
    \includegraphics[width=0.45\textwidth]{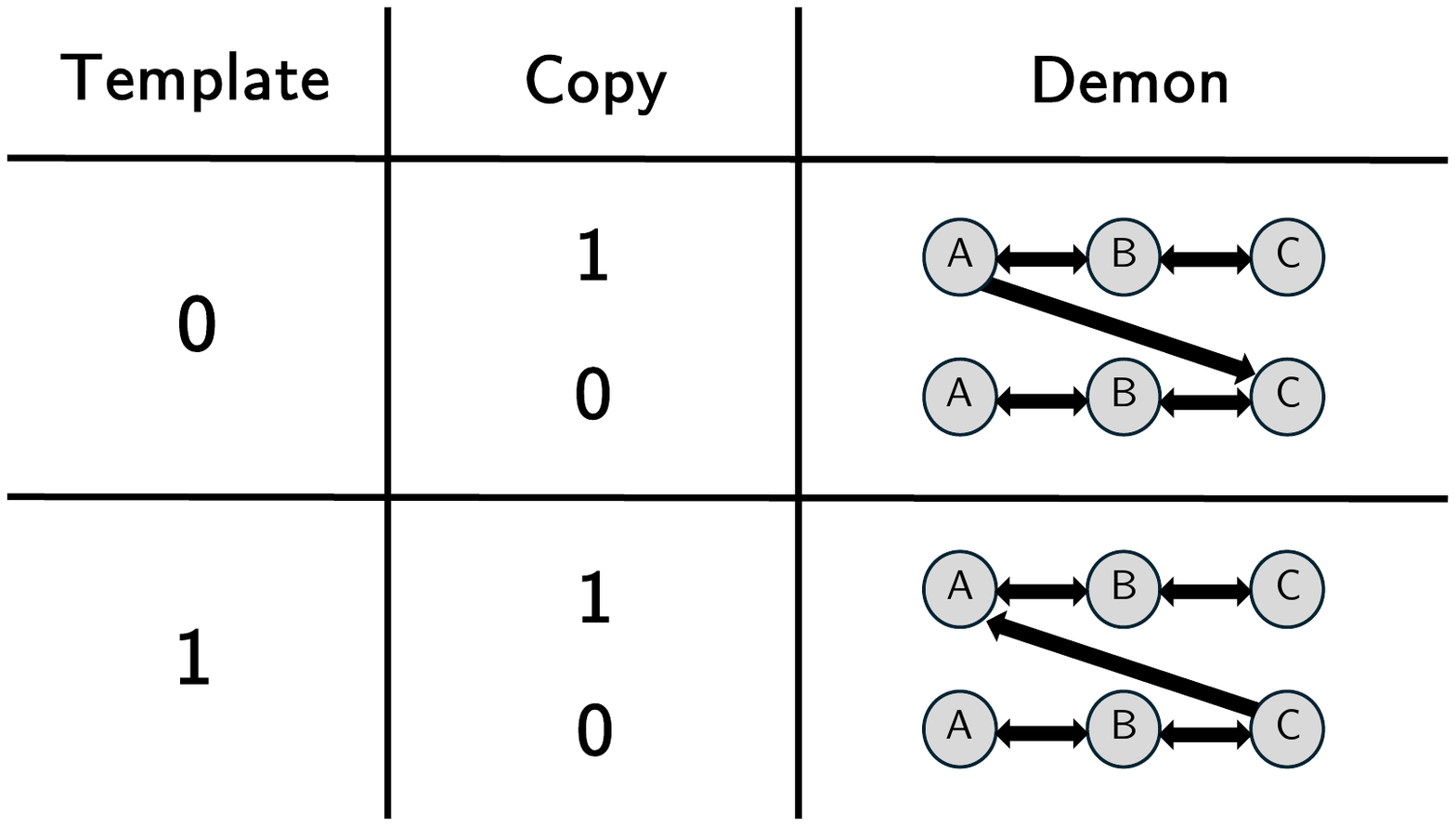}
    \caption{Markov chain representation of information replication process with AMD system. Note the change in directionality of the transitions connecting states A1 and C0 depending on the template bit.}
    \label{fig:RNA}
\end{figure}

In Fig. \ref{fig:RNA}, we provide a visual representation of the rate matrix describing the transcription process.
For the following analysis, we are interested in solving the steady-state dynamics as a function of interaction time $\tau$ and thermodynamic bias $\epsilon$.

\subsection{Analysis of Transcription Fidelity}
\label{sec:fidelities}

We identify two sources of error in the process of information transcription. The first source is in the reading of information, which occurs when the interaction time $\tau$ with each `template' bit is not sufficiently long. The second source is in the \textit{mis}reading of information, which occurs when $|\epsilon| \neq 1$, meaning the optimal (anti) erasure process is obstructed. We analyze the fidelity of information transcription as a function of both parameters in order to identify the optimal parameters for optimal transcription fidelity.

The solution to Eq.~\eqref{eqn:master-equation} is formally given by,
\begin{equation}
    \label{eqn:ME-solution}
    \textbf{P}(\tau)= \textbf{P}_{0}\,  e^{\textbf{R} \tau}.
\end{equation}
Hence, we can interpret the resulting dynamics in terms of interaction intervals per transcribed bit.

We computed the stationary state as a function of $\epsilon$ and $\tau$ using the Gillespie algorithm \cite{gillespie}, which generates individual stochastic trajectories governed by the rate matrix $\mathbf{R}$. The simulation used an equal distribution of $0$ and $1$ bits in the `template' bitstream, for which single-run trajectories of the corresponding `copy' bitstreams were computed. By averaging over an ensemble of these template–copy trajectories, one obtains the steady-state dynamics of the `copy' bitstream, yielding its equilibrium distribution as a function of $\epsilon$ and $\tau$. The transcription fidelity, defined in Eq.~\eqref{eqn:fidelity}, is then computed by simply counting the fraction of bits in the `copy' bitstream that match those in the corresponding `template' bitstream.

Thus, we define the \textit{fidelity} of information transcription $F(\epsilon,\tau)$ as the ensemble-averaged number of total correct `copy' bits $N_{C}$ per the number of total `template' bits $N_{T}$,
\begin{equation}
F(\epsilon, \tau) \equiv \frac{\langle N_{C}(\epsilon,\tau) \rangle}{N_{T}}
\label{eqn:fidelity}
\end{equation}
which we calculate by solving Eq. \eqref{eqn:ME-solution} for the stationary output bit statistics. The bit statistics follow directly from the $12$–state probability vector, grouped into four alternating $(0,1,0,1)$ blocks, where each block contains the $\{\mathrm{A},\mathrm{B},\mathrm{C}\}$ states corresponding to the same bit value.

\begin{figure}
    \centering
    \includegraphics[width=0.48\textwidth]{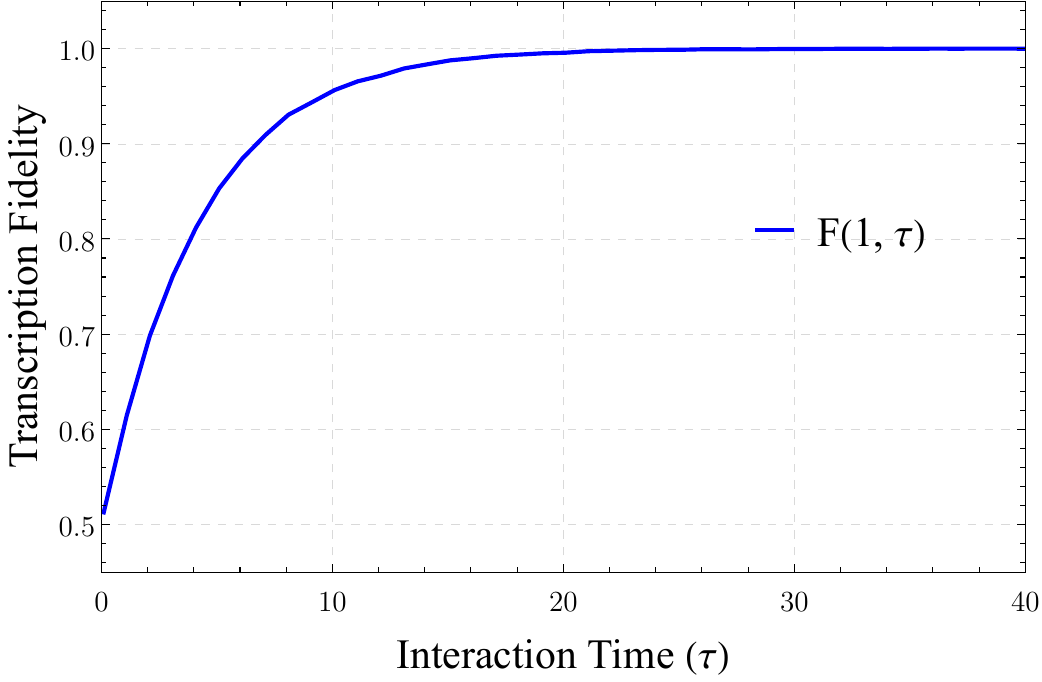}
    \caption{Transcription fidelity $F$ as a function of interaction time $\tau$ with each 'template' bit for 1,000 trajectories. Here, $|\epsilon|$ = 1 to simulate perfect erasure/anti-erasure processes.}
    \label{fig:fidelity-intime}
\end{figure}

To examine the readout error, we computed the fidelity for $|\epsilon| = 1$, which approaches unity as a function of $\tau$ for perfectly accurate transcription. From this, we identify the interaction interval $\tau$ associated with each template–copy bit pair (Fig.~\ref{fig:fidelity-intime}), corresponding to the timescale necessary to minimize intrinsic transcription error.

In Fig. \ref{fig:fidelity-intime}, we observe that $\tau = 30$ is the cutoff point where the transcription error becomes negligibly small, with $F(1, 30) \geq 0.999$. This result can be interpreted to mean that, for an error-free demon–bit system ($|\epsilon| = 1$), the demon must interact with one template-copy bit pair for at least $30$ interaction intervals $\tau$, where for every $10,000$ template bits, only $1$ is wrongly transcribed. Any $\tau \geq 30$ is then guaranteed to have less than or equal to $ 0.01\%$ error in the readout process.

We next evaluated $F(\epsilon,30)$ to confirm the expected limit in which a perfect transcription process arises as $|\epsilon| \rightarrow 1$.
\begin{figure}
    \centering
    \includegraphics[width=0.48\textwidth]{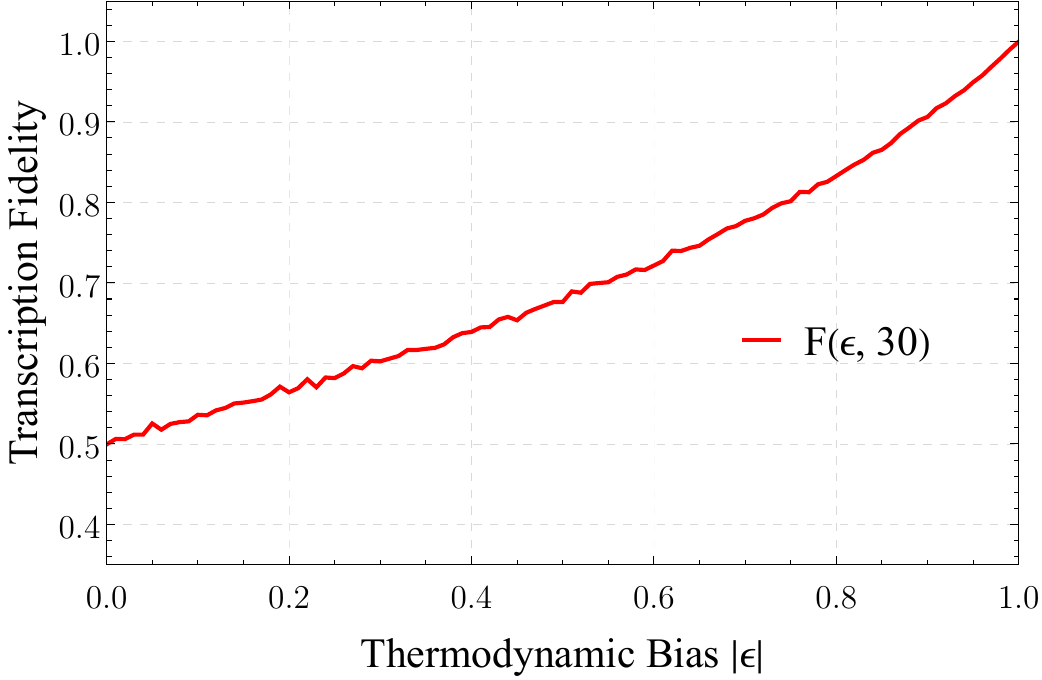}
    \caption{Transcription fidelity $F$ on a lin-log scale as a function of thermodynamic bias $\epsilon$ for 1,000 trajectories. Here, $\tau=30$ to minimize readout error.}
    \label{fig:enter-label}
\end{figure}
Here, we see that fidelity in transcription saturates to $1$ when $|\epsilon| \to 1$, and indicating that the error in the 'copy' bit transcription diminishes to zero, which is consistent with our expectations.

Lastly, we analyzed the mutual information between individual template–copy bit pairs in the steady-state limit as a function of the thermodynamic bias $\epsilon$, where mutual information quantifies the reduction in uncertainty about one random variable given knowledge of another \cite{Shannon1948}. Formally, it is defined in terms of the joint and conditional probabilities as, \begin{equation}
I(X;Y) = \sum_{x,y} p(x,y) \log \!\left[\tfrac{p(x,y)}{p(x)p(y)}\right]\,,
\end{equation}
which can also be interpreted as the difference between the joint entropy and the conditional entropies of the two variables $x$ and $y$. Then, the ensemble-averaged mutual information was computed numerically,
\begin{equation}
    \langle I[N_{T};N_{C}] \rangle
    = \langle H[N_{T},N_{C}] - H[N_{T}|N_{C}] - H[N_{T}|N_{C}] \rangle
    \label{eq:mutualinfo} \ .
\end{equation}

\begin{figure}
    \centering
    \includegraphics[width=0.48\textwidth]{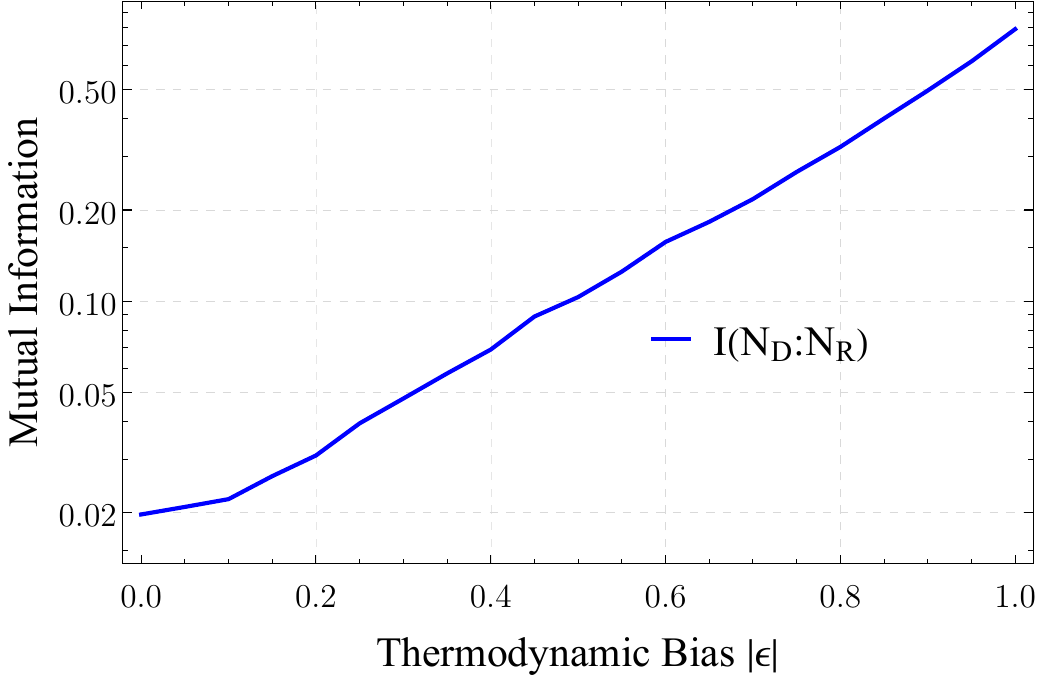}
    \caption{Ensemble-averaged mutual information between template-copy bit pair as a function of thermodynamic bias $\epsilon$ for 1,000 trajectories.}
    \label{fig:mutual-info}
\end{figure}

For clarity, Fig.~\ref{fig:mutual-info} is plotted on a linear–log scale. Along the $y$-axis, the mutual information $I(N_T : N_C) \approx 0.7$ when $|\epsilon| = 1$, corresponding to $\ln 2$. This value represents the maximum possible information that one random variable can provide about another in a two-state system \cite{landauer}. In this regime, the outgoing `copy' bit contains maximal information about the incoming `template' bit, indicating perfectly faithful transcription. Conversely, when $\epsilon \to 0$, the mutual information decreases to $I(N_T : N_C) \approx 0.02$, corresponding to $\sim \ln(1.02)$, which effectively means that measuring the `copy' bit yields almost no information about the original `template' bit. This transition from $I(N_T : N_C) = \ln 2$ to near zero captures the progressive loss of correlation between the template and copy registers, signifying the degradation of information transcription as the system moves from operational to a defective regime.

\section{Time-Inhomogeneous Rates}

If a `healthy' demon is thought to perfectly transcribe information from a template into a blank register, then aging can be understood as the demon’s gradual inability to perform accurate transcription over time. Physically, we know that the death of biological organisms occurs when they thermodynamically equilibrate with their environment \cite{nelson2004biological}. We then interpret the death of the demon-bit system to be when the demon can no longer lift the mass, and the system effectively stalls. 

In the original model, a clockwise transition $C0 \rightarrow A1$ corresponds to the demon lifting the mass while simultaneously writing a bit of information into the bitstream, thereby decreasing the system entropy by $k_B \ln 2$ in accordance with Landauer’s principle \cite{landauer}. Conversely, the reverse transition $A1 \rightarrow C0$ lowers the mass and erases one bit of information. 

Here, we abstract the incoming `template' bit and the outgoing `copy' bit into a single bitstream, such that the demon processes and propagates information through itself. In this representation, we are concerned only with the `copy' register, which records the demon’s transcription activity. To model degradation in transcription fidelity, we introduce a time-dependent bias in the rates of clockwise and counterclockwise transitions, denoted by $\epsilon(t)$.

The time-dependent $6 \times 6$ matrix which generates the dynamics is shown in Eq.~\eqref{eqn:rate-matrix},
\begin{equation}
\textbf{R}(t) = \begin{pmatrix}
-1 & 1 & 0 & 0 & 0 & 0 \\
1 & -2 & 1 & 0 & 0 & 0 \\
0 & 1 & -2+\epsilon(t) & 1+\epsilon(t) & 0 & 0 \\
0 & 0 & 1-\epsilon(t) & -2-\epsilon(t) & 1 & 0 \\
0 & 0 & 0 & 1 & -2 & 1 \\
0 & 0 & 0 & 0 & 1 & -1
\label{eqn:rate-matrix}
\end{pmatrix}.
\end{equation}
The master equation governing the dynamics is given by Eq.~\eqref{eq:TD-ME}, with a time-dependent rate matrix $\mathbf{R}(t)$,
\begin{equation}
    \label{eq:TD-ME}
    \dot{\mathbf{P}} = \mathbf{R}(t) \, \mathbf{P}(t),
\end{equation}
and the corresponding steady-state condition is expressed in Eq.~\eqref{eq:TD-PSS},
\begin{equation}
    \label{eq:TD-PSS}
    \mathbf{R} \, \mathbf{P}_\mathrm{{PSS}} = 0 \ .
\end{equation}
which is now a function of a time-dependent $\epsilon(t)$.

\subsection{Time-Dependent Protocol}

We focus on the steady-state behavior of the composite demon–bit system as it evolves under a time-dependent perturbation to the transition rates. To simulate aging, the time-dependent protocol $\epsilon(t)$ must satisfy several physical and mathematical conditions. Since the model is now abstracted to a single \textit{copy} bitstream initialized to 0s and processed through the demon, aging can be interpreted as the progressive inability to accurately transcribe information into the  copy register. Accordingly, $\epsilon(t)$ should evolve from $+1$ to $0$, representing the transition from a regime in which the demon extracts work from the lifted mass and increases informational order, to one in which it can no longer extract work or write meaningful information, effectively losing memory. In this long-time limit, the demon cannot distinguish between $0$s and $1$s, and the transcription process becomes a random coin flip, representing the system’s informational ``death''. For a time-dependent rate matrix $\mathbf{R}(t)$, the steady state $\mathbf{P}^{(n)}_{\mathrm{PSS}}$ now refers to the \textit{instantaneous} steady state because it changes as a function of $\epsilon(t)$, which is clarified by using a superscript $(n)$.

To capture this behavior, the time-dependent bias $\epsilon(t)$ must satisfy three criteria: (i)  it must remain bounded between $\pm 1$ to satisfy detailed balance and ensure a unique steady state \cite{maxwell, vanKampen2007}; (ii)  it must vary smoothly, quasi-statically and be differentiable (Lipschitz-continuous) so that $\mathbf{P}(t)$ closely follows $\mathbf{P}^{(n)}_{\mathrm{PSS}}$ \cite{Lipschitz1864}; and (iii) it must evolve monotonically from $+1 \to 0$ to simulate the gradual loss of the demon’s ability to transcribe meaningful information properly, corresponding to memory degradation and aging.

\begin{figure}
    \centering
    \includegraphics[width=0.48\textwidth]{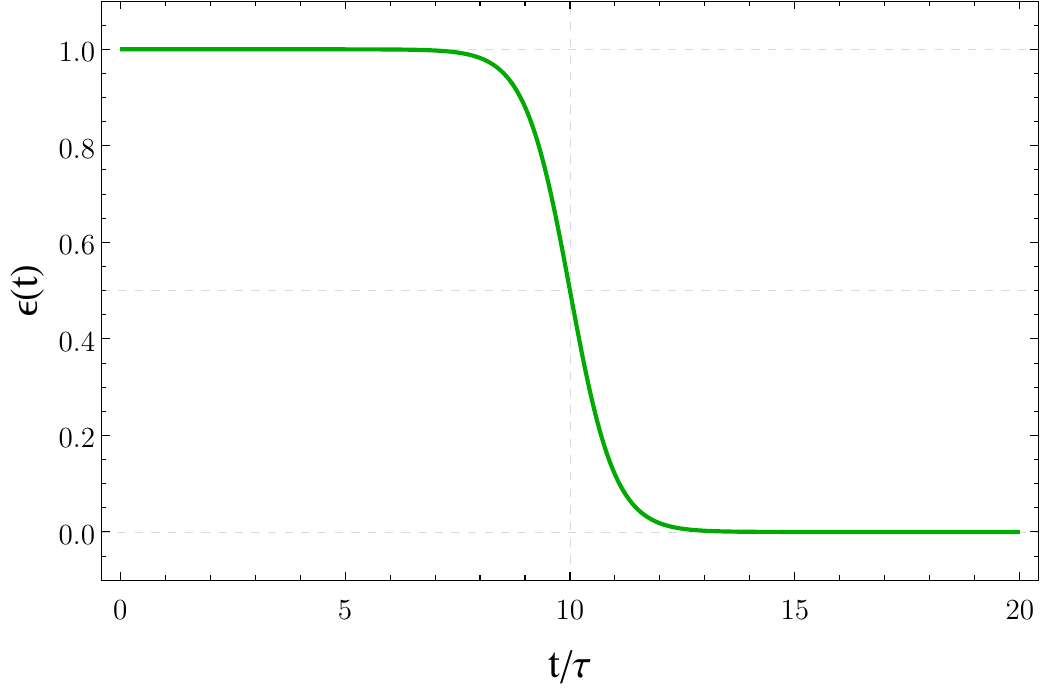}
    \caption{Time-dependent protocol of $\epsilon(t)=-\tanh{\left[(t-t_0)/\tau\right]}$ to simulate aging system.}
    \label{fig:tanhplot}
\end{figure}

We therefore choose the time-dependent protocol to be in the general functional form $\epsilon(t)$ shown in Eq. \eqref{eq:eps-protocol}
\begin{equation}
    \label{eq:eps-protocol}
    \epsilon(t) = A\tanh\!\left(\frac{t-t_0}{\tau}\right) + B,
\end{equation}
and use parameters $A=-1/2$, $B=1/2$, $t_0 = t_{max}/2=300$, and $\tau = 30$. This function provides a smooth and monotonic transition from $+1$ to $0$, consistent with the gradual loss of informational and thermodynamic efficiency described above; however, any suitably bounded and differentiable function could be used to represent alternative driving protocols. The time-dependent protocol is visualized in Fig. \ref{fig:tanhplot}.

To validate convergence of the system to a `target' state, we evaluate the distance between the instantaneous probability vector $\mathbf{P}(t)$ and its corresponding instantaneous steady state $\mathbf{P}^{(n)}_{\mathrm{PSS}}$, given by the right eigenvector of $\mathbf{R}(t)$ associated with the null eigenvalue \cite{Risken1996}. In the adiabatic limit, this distance approaches zero,
\begin{equation}
\label{eq:contraction}
    \lim_{t \to \infty} \left\| \mathbf{P}(t) - \mathbf{P}^{(n)}_{\mathrm{PSS}} \right\| = 0 \ .
\end{equation}
See App. \ref{app:proof} for an explicit proof of this statement.

When the rate matrix $\mathbf{R}(t)$ is Lipschitz differentiable \cite{Chetrite2018}, meaning its rate of change is bounded and varies smoothly in time, both $\mathbf{P}(t)$ and its instantaneous steady state $\mathbf{P}^{(n)}_{\mathrm{PSS}}$ vary smoothly in time. If the rate of change of $\mathbf{R}(t)$ is small compared to the spectral gap separating the null eigenvalue from the rest of the spectrum, the dynamics remain adiabatic. Thus, the system follows a quasi-static trajectory on the thermodynamic manifold defined by $\epsilon(t)$, and $\mathbf{P}(t)$ stays arbitrarily close to $\mathbf{P}^{(n)}_{\mathrm{PSS}}$ throughout the evolution.

\begin{figure}
    \centering
    \includegraphics[width=0.48\textwidth]{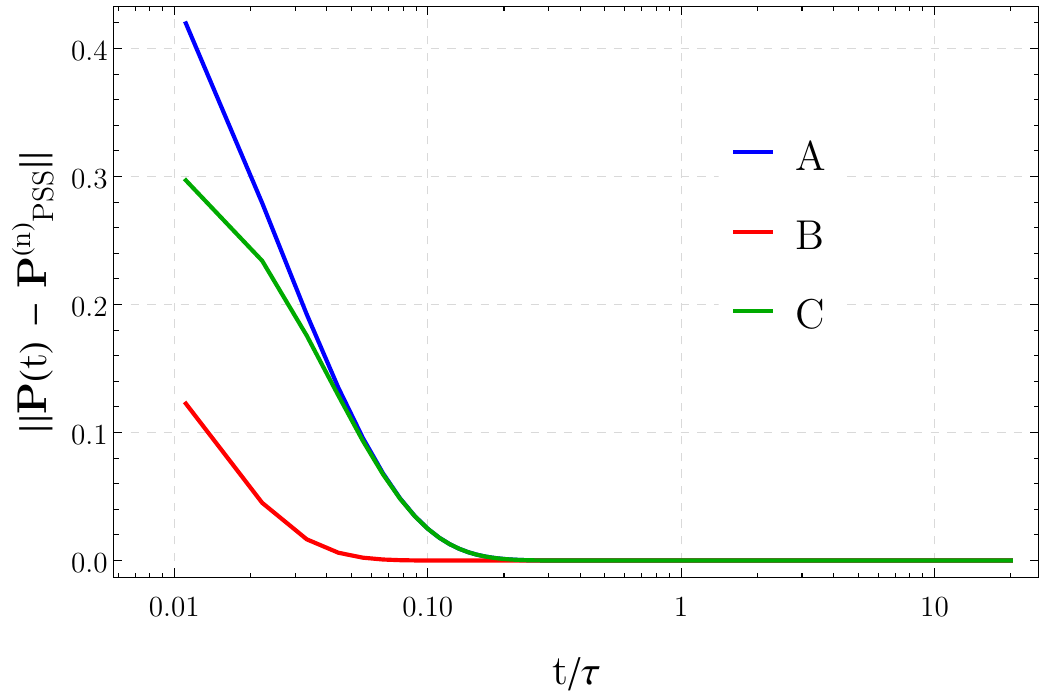}
    \caption{Log-linear scale convergence of distance between instantaneous state and instantaneous steady state for demon state $\{\mathrm{A,B,C}\}$ for time-dependent protocol $\epsilon(t)$.}
    \label{fig:pss-converge}
\end{figure}

We numerically verified this behavior in Fig. \ref{fig:pss-converge}, which shows the normed distance between $\mathbf{P}(t)$ and $\mathbf{P}^{(n)}_{\mathrm{PSS}}$ converging to zero as $\epsilon(t)$ evolves smoothly.
We observe that the normed distance between $\mathbf{P}(t)$ and $\mathbf{P}^{(n)}_{\mathrm{PSS}}$ vanishes over time, guaranteeing the system's convergence to the instantaneous steady state.

\subsection{Notions of Aging I: Extractable Work}

\begin{figure}
    \centering
    \includegraphics[width=0.48\textwidth]{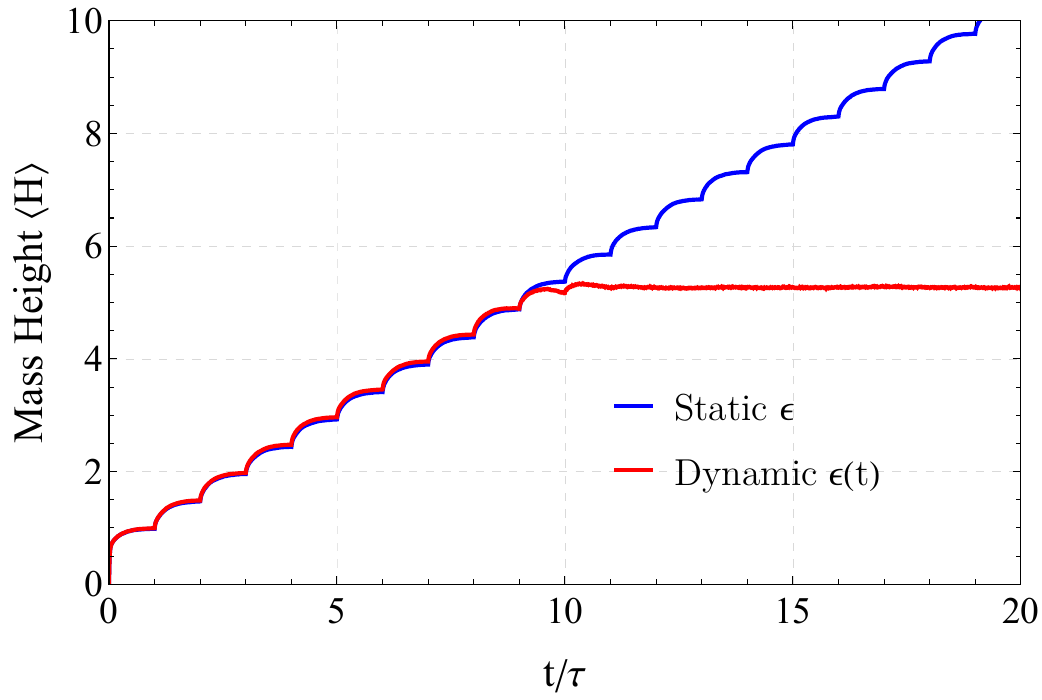}
    \caption{Comparison of mean mass height for static vs. time-dependent $\epsilon(t)$ protocol for 1,000 trajectories. A clear performance drop is observed as $\epsilon(t)$ decreases.}
    \label{fig:mass_height_comparison}
\end{figure}

We compared the change in the average height of the suspended mass under both static and time-dependent $\epsilon$ protocols to quantify the degradation in the demon’s work extraction capacity (Fig. \ref{fig:mass_height_comparison}). When the demon writes one bit of information ($0 \rightarrow 1$), it extracts an amount of work \cite{landauer},
\begin{equation}
    W_{+} = k_{\mathrm{B}} T \ln 2
\end{equation}
lifting the mass by one unit height. Conversely, it has been proven that when a bit is erased ($1 \rightarrow 0$), the demon dissipates the same amount of energy back into the thermal reservoir \cite{Boyd2016},
\begin{equation}
    W_{-} = -k_{\mathrm{B}} T \ln 2
\end{equation}
lowering the mass by one unit height. The total work performed along a single trajectory of duration $t_{\max}$ can therefore be expressed as
\begin{equation}
    W(t_{\max}) = k_{\mathrm{B}} T \ln 2 \sum_{i=1}^{N_{\mathrm{bits}}} \sigma_i(\tau_i),
\end{equation}
where $\sigma_i = +1$ for a written bit ($0 \rightarrow 1$) and $\sigma_i = -1$ for an erased bit ($1 \rightarrow 0$). Additionally, $N_{\mathrm{bits}} = \frac{t_{\max}}{\tau}$ denotes the total number of template–copy interactions within a trajectory.

To compute the ensemble-averaged behavior, both the static rate matrix $\mathbf{R}$ and the time-dependent matrix $\mathbf{R}(t)$ are evolved over identical durations $t_{\max}$, with the bit value recorded after each interaction interval $\tau_i$. The mean output bit value,
\begin{equation}
    \langle b_{\text{out}} \rangle = \frac{1}{N_{\mathrm{bits}}} \sum_{i=1}^{N_{\mathrm{bits}}} \sum_{j} b_j P_j(\tau_i)
\end{equation}
is proportional to the average mass height and quantifies the effective extractable work per bit.

The ensemble-averaged work output is therefore given by,
\begin{equation}
    \langle W(t_{\max}) \rangle = k_{\mathrm{B}} T \ln 2 \, N_{\mathrm{bits}} \, \langle b_{\text{out}} \rangle
\end{equation}
which provides a direct measure of the demon’s ability to convert information into mechanical work under both static and time-dependent $\epsilon$ protocols. The demon-bit system behavior for both protocols is shown in Fig. \ref{fig:mass_height_comparison}.

As expected, under the time-dependent bias $\epsilon(t)$ the output distribution gradually shifts into a regime where the demon no longer transcribes meaningful information, but instead effectively performs a coin flip when selecting which bit to write. In this limit, the net information written into the `copy' stream vanishes on average, and the demon’s information-processing functionality effectively stalls. This loss of transcription fidelity and associated work capacity culminates as $\epsilon(t)\to 0$, driving the system into a steady, ``vegetative'' state in which the average mass height saturates and no further useful work can be extracted.

\subsection{Notions of Aging II: Entropy Production} 

We finally evaluate the total entropy production rate, $\dot{\Sigma}_{\mathrm{tot}}(t)$, for both the time-independent and time-dependent protocols to quantify how temporal driving modifies dissipation as the demon gradually loses its mechanical and informational function. 

In the static $\epsilon$ case, the dynamics are generated by a time-independent rate matrix $R_{ij}$, so the demon-bit system converges to the nonequilibrium periodic steady state (NESS), $\mathbf{P}_{\mathrm{PSS}}$, which is characterized by stationary probability currents. During the initial relaxation toward $\mathbf{P}_{\mathrm{PSS}}$ there is the transient \textit{adiabatic} entropy production rate, $\dot{\Sigma}_{\mathrm{A}}(t)$, associated with the contraction of the evolving distribution $\mathbf{P}(t)$ to the steady state. Once the NESS is reached, $P_i(t) \to \pi_i$ and the adiabatic rate vanishes, so that the corresponding cumulative adiabatic entropy production saturates to a constant value. Maintaining the NESS, however, requires continual dissipation in the form of the \textit{non-adiabatic} entropy production, $\dot{\Sigma}_{\mathrm{NA}}(t)$, which is associated with sustaining the non-equilibrium currents.

In contrast, under the dynamic $\epsilon(t)$ protocol, the rate matrix $R_{ij}(t)$ and the instantaneous `steady' state $\boldsymbol{\pi}^{(n)}\equiv\mathbf{P}^{(n)}_{\mathrm{PSS}}$ are both time dependent. The dynamics can then be viewed as a competition between the contraction toward $\mathbf{P}^{(n)}_{\mathrm{PSS}}$ and the ``motion'' of the target state itself, as analyzed earlier. Even when the demon has effectively lost information-processing functionality, such as when $\epsilon(t)\to 0$ so that the outgoing bit stream approaches an equal mixture of $0$s and $1$s, the protocol remains a driven, nonequilibrium process as long as the rates are changing and the actual distribution $\mathbf{P}(t)$ lags behind $\mathbf{P}^{(n)}_{\mathrm{PSS}}$. In this regime, the adiabatic contribution $\dot{\Sigma}_{\mathrm{A}}(t)$ stagnates as the instantaneous state contracts to the instantaneous steady state, while the non-adiabatic term $\dot{\Sigma}_{\mathrm{NA}}(t)$ continues to generate dissipation due to this persistent lag. 

We quantify these statements using the standard expression for the total entropy production rate \cite{Esposito2010PRE},
\begin{equation}
\dot{\Sigma}_{\mathrm{tot}}(t)
= \sum_{i<j}
\bigl(P_i(t) R_{ij}(t) - P_j(t) R_{ji}(t)\bigr)
\ln\!\left(\frac{P_i(t)\,R_{ij}(t)}{P_j(t)\,R_{ji}(t)}\right) ,
\label{eq:sigma-total}
\end{equation}
which can be decomposed into the sum of the instantaneous adiabatic and non-adiabatic entropy production rates \cite{Esposito2010PRE},
\begin{equation}
    \dot{\Sigma}_{\mathrm{tot}}(t) = \dot{\Sigma}_{\mathrm{A}}(t) + \dot{\Sigma}_{\mathrm{NA}}(t),
\end{equation}
with
\begin{equation}
\dot{\Sigma}_{\mathrm{A}}(t)
= \sum_{i<j}
\bigl(P_i(t) R_{ij}(t) - P_j(t) R_{ji}(t)\bigr)
\ln\!\left(\frac{\pi_i(t)\,R_{ij}(t)}{\pi_j(t)\,R_{ji}(t)}\right),
\label{eq:AEP}
\end{equation}
and
\begin{equation}
\dot{\Sigma}_{\mathrm{NA}}(t)
= \sum_{i<j}
\bigl(P_i(t) R_{ij}(t) - P_j(t) R_{ji}(t)\bigr)
\ln\!\left(\frac{P_i(t)\,\pi_j(t)}{P_j(t)\,\pi_i(t)}\right),
\label{eq:NAEP}
\end{equation}
where $\pi_i(t)$ are the components of the instantaneous steady-state vector, $R_{ij}(t)$ are the elements of the rate matrix, and $P_i(t)$ are the components of the instantaneous state vector. 

From Eqs.~\eqref{eq:AEP} and \eqref{eq:NAEP} it is clear that when the steady-state distribution is reached $P_i(t \to \infty)=\pi_i$ and detailed balance holds $R_{ij}\,\pi_j = R_{ji}\,\pi_i$, the adiabatic entropy production rate $\dot{\Sigma}_{\mathrm{A}}(t)$ vanishes and the associated cumulative contribution reaches a constant value, leaving only remaining $\dot{\Sigma}_{\mathrm{NA}}(t)$ due to the irreversibility of the process.

We obtain the cumulative entropy production by numerically integrating over the interval $[0,t]$ using a left-Riemann approximation and summing the two contributions,
\begin{equation}
    \Sigma(t) = \int_{0}^{t}\dot{\Sigma}_{\mathrm{A}}(t')\,dt' + \int_{0}^{t}\dot{\Sigma}_{\mathrm{NA}}(t')\,dt' ,
\end{equation}
with time step $dt' = 0.01$.

\begin{figure}
    \centering
    \includegraphics[width=0.48\textwidth]{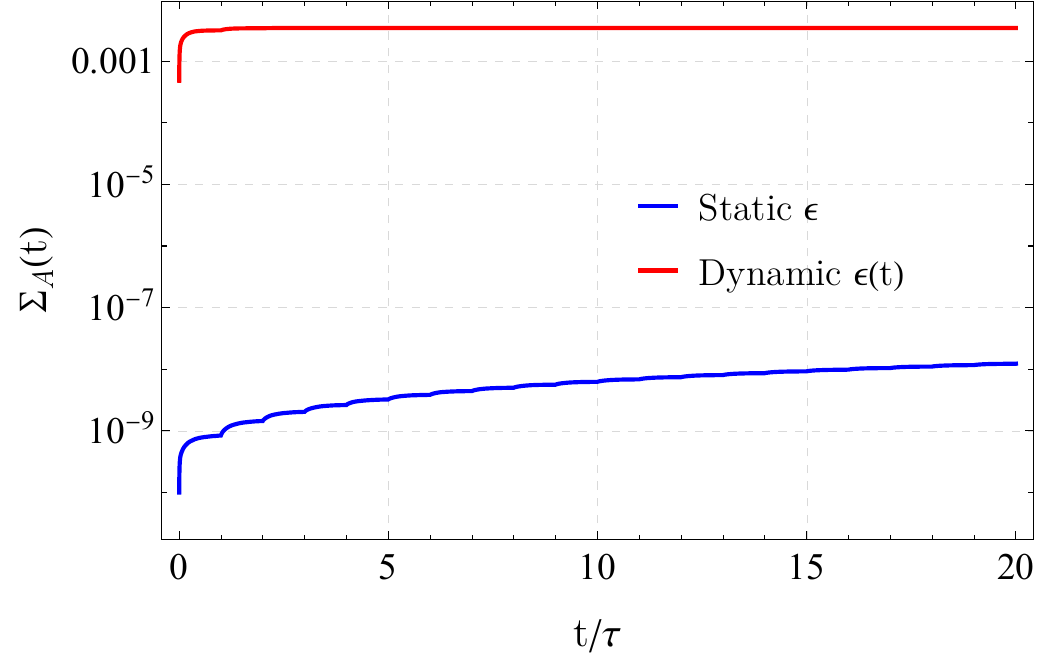}
    \caption{Cumulative adiabatic entropy production for $1{,}000$ realizations of the static $\epsilon$ and dynamic $\epsilon(t)$ protocols.}
    \label{fig:adiabaticentropy}
\end{figure}

For both the time-independent $\epsilon$ and time-dependent $\epsilon(t)$ cases in Fig.~\ref{fig:adiabaticentropy}, the cumulative adiabatic entropy production rapidly levels off, indicating that the adiabatic contribution is negligible beyond an initial transient and that the system is effectively initialized close to its instantaneous steady state. $\Sigma_A(t)$ being larger for $\epsilon(t)$ is due to $\mathbf{P}(t)$ having to `land' in the time-dependent instantaneous steady state.

\begin{figure}
    \centering
    \includegraphics[width=0.48\textwidth]{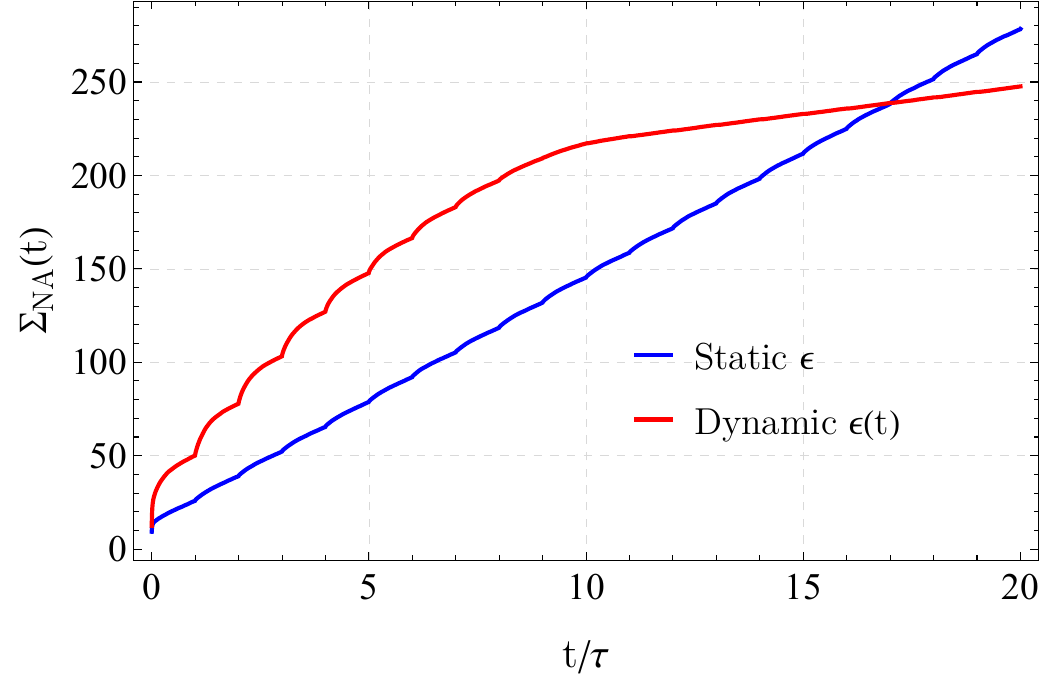}
    \caption{Cumulative non-adiabatic entropy production for $1{,}000$ realizations of the static $\epsilon$ and dynamic $\epsilon(t)$ protocols.}
    \label{fig:nonadiabaticentropy}
\end{figure}

For the non-adiabatic contributions in Fig.~\ref{fig:nonadiabaticentropy}, the two protocols display more distinct behavior. In the static $\epsilon$ case, repeated bit resetting due to the moving bitstream keep the non-adiabatic rate finite, so the cumulative $\Sigma_{\mathrm{NA}}(t)$ grows approximately linearly in time, reflecting the steady dissipation required to maintain the demon’s ability to convert thermal fluctuations into ordered information or useful work. Under the time-dependent $\epsilon(t)$ protocol, $\Sigma_{\mathrm{NA}}(t)$ initially grows more rapidly than in the static case due to the combined cost of contracting toward and simultaneously `chasing' the moving target state $\mathbf{P}^{(n)}_{\mathrm{PSS}}$. However, as $\epsilon(t)\to 0$ the outgoing bit stream becomes unbiased and the demon ceases to produce ordered information, so the cumulative non-adiabatic entropy production continues to increase only slowly and ultimately remains below that of the static protocol.

\begin{figure}
    \centering
    \includegraphics[width=0.48\textwidth]{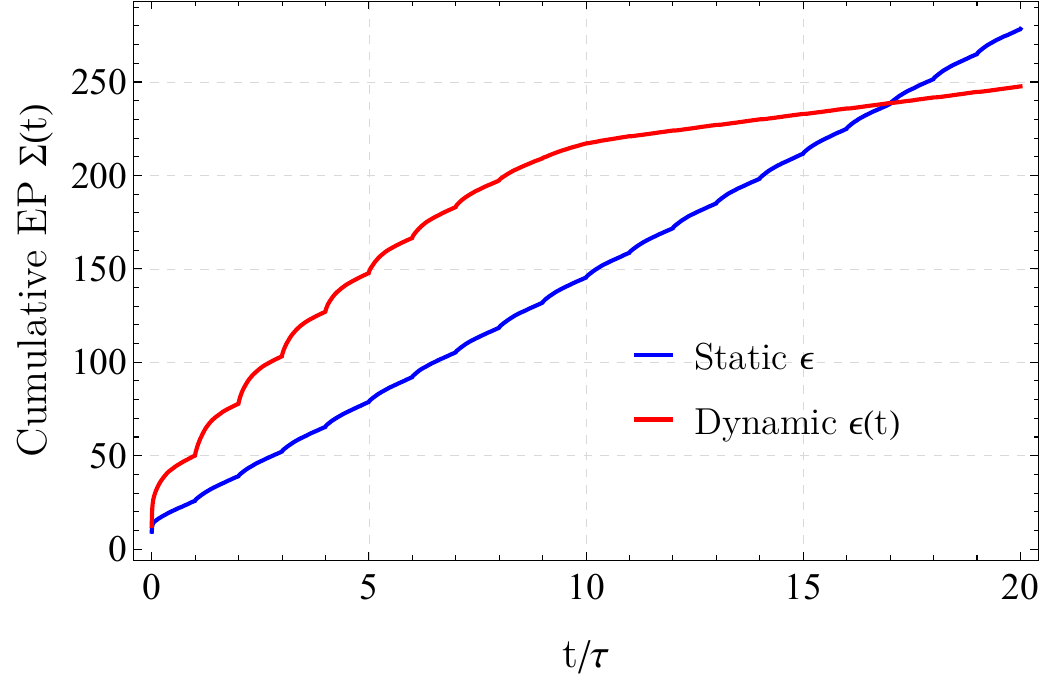}
    \caption{Cumulative total entropy production for $1{,}000$ realizations of the static $\epsilon$ and dynamic $\epsilon(t)$ protocols.}
    \label{fig:entropy}
\end{figure}

Even after the demon has lost its capacity to perform useful work, the process remains genuinely nonequilibrium for as long as $\mathbf{P}(t)$ lags behind the time-dependent target $\mathbf{P}^{(n)}_{\mathrm{PSS}}$, and the residual non-adiabatic dissipation can be interpreted as the additional thermodynamic cost associated with the final stages of relaxation after the decline in the demon's functionality. Consequently, at late times the cumulative total entropy production under the static $\epsilon$ protocol exceeds that of the aging protocol $\epsilon(t)$, capturing both the extra dissipation accrued as the system lags behind a functional target state and the eventual leveling off of entropy production as the demon loses its ability to process information accurately. From a biological standpoint, this behavior is consistent with the picture that aging cellular machinery gradually loses the ability to maintain ordered, functional states. In our model, the limit $\epsilon(t)\to 0$ corresponds to the demon no longer biasing the outgoing bit stream, effectively ``forgetting'' how to correctly transcribe meaningful information and instead producing an unbiased output akin to a coin flip. Thus, the stagnation of the non-adiabatic entropy production rate signals equilibration with the environment which we interpret as the ``death'' of the system, while the remaining slow increase in non-adiabatic entropy production reflects the protocol-induced lag behind $\mathbf{P}^{(n)}_{\mathrm{PSS}}$ that can be heuristically associated with post-mortem decomposition and the breakdown of cellular components even after information processing has ceased.

\section{Concluding Remarks}

In this paper, we have developed a minimal, time-dependent model of an autonomous Maxwell’s demon to study biological aging as a gradual loss of information-processing fidelity. By embedding a time-dependent thermodynamic bias $\epsilon(t)$ into the demon’s transition rates, we simulated the transition from a functional regime, where work is extracted and information is accurately transcribed, to a degraded regime where transcription halts and no useful work can be performed. In doing so, we established a general recipe for modeling the decay in biological processes using stochastic thermodynamics and Maxwellian information ratchets: define the biological function as information transfer, encode loss of fidelity as a time-dependent bias, and track the evolution of entropy production, mutual information, and work extraction.

Our analysis shows that when the generator $\mathbf{R}(t)$ is Lipschitz-differentiable with a finite spectral gap, the system follows its instantaneous steady state closely, allowing the progressive decay of informational and thermodynamic quantities to be captured in closed form. As $\epsilon(t) \to 0$, mutual information and extractable work both decrease, and the cumulative dissipation of the system increases, reflecting the demon’s loss of functionality. This alignment between thermodynamic irreversibility and information degradation provides a quantitative link between microscopic aging, energy dissipation, and fidelity loss.

These results place Maxwellian information ratchets within a broader biological context, where the same principles that govern molecular machines, such as RNA polymerase and ATP synthase, also govern their decline. By extending stochastic thermodynamics to time-inhomogeneous dynamics, this framework offers a compact framework to understand the relevance of information fidelity and thermodynamic efficiency in living systems.

In the future, error-correcting mechanisms could be incorporated to study how cells mitigate transcriptional errors \cite{RNAerrorcorrect}. Another caveat that could be explored in the future is the cutoff in transcription fidelity as temperature becomes too extreme for the cell to endure. This could open the door for the simulation of cancer development, epigenetic diseases, and more. The framework may also be adapted to include errors arising in other processes such as mRNA translation. From a computational perspective, it offers a way to benchmark numerical solvers for biologically relevant stochastic differential equations \cite{Gillespie2001JCP,Voraldo2002JTB,Anderson2007JCP}. Treating the driving parameter $\epsilon(t)$ as a genuinely stochastic variable would also allow the model to account for environmental variability across cell populations. 

To summarize, this paper provides a simplified framework for analyzing biological aging through the lens of information theory and thermodynamics, and benchmarks future work for connecting stochastic thermodynamics with biological processes in microscopic systems.

\acknowledgments{Student support provided by an NIGMS Graduate Research Training Initiative for Student Enhancement (G-RISE) Grant (T32-GM144876). S.D. acknowledges support from the John Templeton Foundation under Grant No. 63626.}

\appendix

\section{Read-Write Fidelity \label{sec:appendix}}

In this appendix, we justify the use of the same coupling rate to both bit streams. To this end, imagine a scenario where DNA polymerase might couple more strongly to either the DNA stream or the RNA stream, potentially affecting the DNA transcription fidelity in the process. 

We consider the read-write fidelity $F(\epsilon_R,\epsilon_W)$, where we can conceptualize the transcription process being only as efficient as the worst performer. This read-write fidelity is calculated by first splitting each interaction window $\tau$ into two \textit{equal} parts, where first the demon reads the incoming bit from the `template' register, and then writes that information into the `copy' register. The read-write fidelity is thus defined as,
\begin{equation}
    F(\epsilon_R,\epsilon_W) = \min \{ F(\epsilon_R,30), \, F(\epsilon_W,30) \}
\end{equation}
where $\tau =30$ to reflect the optimal interaction time for the template-copy bit pair. The fidelity is evaluated the same way as in Sec.~\ref{sec:fidelities}. 

\begin{figure}
    \centering
    \includegraphics[width=0.48\textwidth]{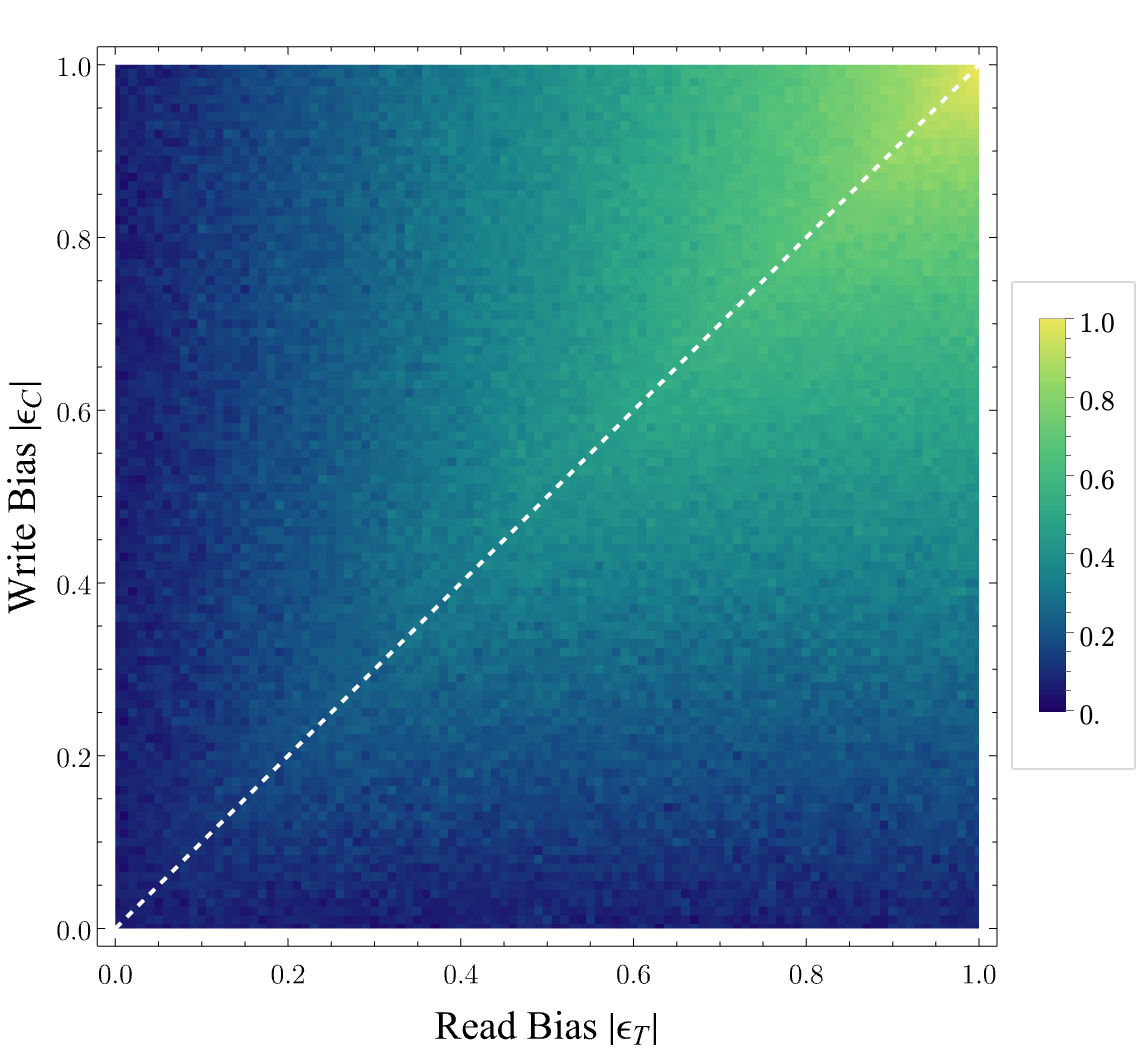}
    \caption{Read-write fidelity $F(\epsilon_R,\epsilon_W)$ for 100 trajectories, white dotted line denoting when $\epsilon_R=\epsilon_W$, exhibiting linear growth.}
    \label{fig:two-eps-fidelity}
\end{figure}

In Fig.~\ref{fig:two-eps-fidelity}, we observe that the fidelity grows linearly between $\epsilon_R$ and $\epsilon_W$, saturating to unity when $|\epsilon_R|=|\epsilon_W|=1$.

\section{Contraction to Instantaneous Steady State \label{app:proof}}

Finally, we provide a proof of Eq.~\eqref{eq:contraction}. To this end, we have to show that at each step, the probability vector $\mathbf{P}_n(t)$ contracts toward its instantaneous steady state $\mathbf{P}^{(n)}_{\mathrm{ss}}$ more than $\mathbf{P}^{(n)}_{\mathrm{ss}}$ itself shifts between steps. This supports convergence to the steady state over long timescales. 

Let the transition probability matrix be diagonalizable as $\mathbf{Q}_1 = \mathbf{U}_1 \mathbf{D}_1 \mathbf{U}_1^{-1}$, where $\mathbf{D}_1$ contains eigenvalues $\lambda_i$,
\begin{equation}
    \mathbf{D}_1 = \begin{bmatrix}
        1 & 0 & 0 \\
        0 & \lambda_2 & 0 \\
        0 & 0 & \lambda_3
    \end{bmatrix}
    ,
\end{equation}
labeled in descending order.
Here, $\lambda_1 = 1$ corresponds to the steady state, and the remaining eigenvalues $|\lambda_i| < 1$ for $i > 1$ govern the decay rate of inhomogeneities present in the instantaneous probability vector $\mathbf{P}_{n}(t)$. For simplicity, we set $\lambda_2 = \lambda_3$ in order to derive a lower bound on the rate of convergence.

Looking at just one time step, the initial distribution can be decomposed as
\begin{equation}
    \mathbf{P}_1 = \mathbf{P}^{(1)}_{\mathrm{ss}} + \beta \mathbf{B} + \gamma \mathbf{C},
\end{equation}
where $\mathbf{B}$ and $\mathbf{C}$ are “error” vectors. Applying $\mathbf{D}_1$ yields
\begin{equation}
    \mathbf{D}_1 \mathbf{P}_1 = \mathbf{P}^{(1)}_{\mathrm{ss}} + \lambda_2 \bigl(\beta \mathbf{B} + \gamma \mathbf{C}\bigr)\,.
\end{equation}
Thus, the distance to steady state contracts as
\begin{equation}
    \bigl\|\mathbf{D}_1 \mathbf{P}_1 - \mathbf{P}^{(1)}_{\mathrm{ss}}\bigr\| = \lambda_2 \,\|\beta \mathbf{B} + \gamma \mathbf{C}\|.
\end{equation}
Comparing to the original distance, we have
\begin{equation}
    \bigl\|\mathbf{P}_1 - \mathbf{P}^{(1)}_{\mathrm{ss}}\bigr\| = \|\beta \mathbf{B} + \gamma \mathbf{C}\|.
\end{equation}
Thus, contraction occurs whenever $\lambda_2 < 1$,
\begin{equation}
    \bigl\|\mathbf{D}_1 \mathbf{P}_1 - \mathbf{P}^{(1)}_{\mathrm{ss}}\bigr\| \leq \bigl\|\mathbf{P}_1 - \mathbf{P}^{(1)}_{\mathrm{ss}}\bigr\|.
\end{equation}
In the edge case $\lambda_2 = 1$, no contraction occurs. For primitive stochastic matrices, the Perron–Frobenius theorem guarantees $\lambda_2 < 1$, ensuring asymptotic convergence \cite{PFTheorem, Edelman2007}.

\subsubsection*{Distance to a Moving Target State}

When $\epsilon(t)$ varies, the steady state shifts, 
$\mathbf{P}^{(n)}_{\mathrm{ss}} \to \mathbf{P}^{(n+1)}_{\mathrm{ss}}$. To analyze this, express the change in target state as
\begin{equation}
    \mathbf{P}^{(n+1)}_{\mathrm{ss}} = \mathbf{P}^{(n)}_{\mathrm{ss}} + c_2 \mathbf{B} + c_3 \mathbf{C}.
\end{equation}
The new distance from $\mathbf{P}_n$ to $\mathbf{P}^{(n+1)}_{\mathrm{ss}}$ becomes
\begin{align}
    \bigl\|\mathbf{P}_n - \mathbf{P}^{(n+1)}_{\mathrm{ss}}\bigr\| 
    &= \bigl\|\beta \mathbf{B} + \gamma \mathbf{C} - c_2 \mathbf{B} - c_3 \mathbf{C}\bigr\| \\
    &= \bigl\|(\beta - c_2)\mathbf{B} + (\gamma - c_3)\mathbf{C}\bigr\|.
\end{align}

Leveraging the triangle inequality, we obtain
\begin{equation}
    \bigl\|\mathbf{P}_n - \mathbf{P}^{(n+1)}_{\mathrm{ss}}\bigr\| 
    \leq \|\beta \mathbf{B} + \gamma \mathbf{C}\| + \|c_2 \mathbf{B} + c_3 \mathbf{C}\|.
\end{equation}
Identifying the terms,
\begin{align}
    \|\beta \mathbf{B} + \gamma \mathbf{C}\| &= \bigl\|\mathbf{P}_n - \mathbf{P}^{(n)}_{\mathrm{ss}}\bigr\|, \\
    \|c_2 \mathbf{B} + c_3 \mathbf{C}\| &= \bigl\|\mathbf{P}^{(n+1)}_{\mathrm{ss}} - \mathbf{P}^{(n)}_{\mathrm{ss}}\bigr\|,
\end{align}
we have
\begin{equation}
    \bigl\|\mathbf{P}_n - \mathbf{P}^{(n+1)}_{\mathrm{ss}}\bigr\| 
    \leq \bigl\|\mathbf{P}_n - \mathbf{P}^{(n)}_{\mathrm{ss}}\bigr\| 
    + \bigl\|\mathbf{P}^{(n+1)}_{\mathrm{ss}} - \mathbf{P}^{(n)}_{\mathrm{ss}}\bigr\|.
\end{equation}

Now, including the contraction proven above, we can write
\begin{equation}
    \bigl\|\mathbf{Q}_n \mathbf{P}_n - \mathbf{P}^{(n)}_{\mathrm{ss}}\bigr\| 
    + \bigl\|\mathbf{P}^{(n+1)}_{\mathrm{ss}} - \mathbf{P}^{(n)}_{\mathrm{ss}}\bigr\| 
    \leq \bigl\|\mathbf{P}_n - \mathbf{P}^{(n)}_{\mathrm{ss}}\bigr\|.
\end{equation}
This inequality ensures that the contraction toward $\mathbf{P}^{(n)}_{\mathrm{ss}}$ dominates over the shift in target state, allowing the system to track the moving target state under the slowly varying protocol $\epsilon(t)$.

\bibliography{new}

\end{document}